# Upper critical field anisotropy in $BaFe_{2-x}Co_xAs_2$ single crystals synthesized without flux

K. Vinod, A. T. Satya, Shilpam Sharma, C. S. Sundar, and A. Bharathi*

*Material Science Group, Indira Gandhi Center for Atomic Research, Kalpakkam 603102 India*

**Abstract**

The upper critical field was measured upto 12 T for three $BaFe_{2-x}Co_xAs_2$ single crystals with estimated Co concentrations of $x = 0.082$, $x = 0.117$ and $x = 0.143$. $H_{C2}$ versus temperature was measured from temperature dependent resistivity, for various applied magnetic fields, H || ab and H || c. The $[dH_{C2}/dT]_{T=Tc}$, normalized with the corresponding $T_C$, decreases with increasing Co content, for both directions. The anisotropy $\gamma$ defined as $H_{C2}$ || ab / $H_{C2}$ || c shows a distinct increase with Co content, and its temperature dependence shows a peak close to the $T_C$. Magneto transport measurements, in the spin density wave regions, showed significant negative MR for H || ab and positive MR of H || c in the $x = 0.082$ sample. The implications of these results are discussed.

*Corresponding author

A. Bharathi

Condensed Matter Physics Division,

Materials Science Group,

Indira Gandhi Center for Atomic Research,

Kalpakkam, 603102 India.

Tel: 91 44 27480081, Fax: 91 44 27480081

E-mail: bharathi@igcar.gov.in

The $BaFe_{2-x}Co_xAs_2$ system has been widely investigated, resulting in reproducible structural/magnetic and superconducting phase diagrams.[1-3] As Co dopes at the Fe site, structural phase transition temperature ($T_{SPT}$) and spin density wave transition temperature ($T_{SDW}$) that occur at the same temperature in the pristine sample, separate and are suppressed by roughly 15 K per atomic percent of Co. For Co substitution of ~ 4% superconductivity sets in at low temperature of ~ 7 K. $T_C$ increases and reaches a maximum at an optimal doping of ~ 10% Co which is followed by small decrease, resulting in a dome like $T_C - x$ phase diagram. [1-3] In the under-doped region the samples exhibit structural, magnetic and superconducting transitions.[1,3] This regime that exhibits co-existence of SDW and superconducting states, is associated with unusual properties, like in-plane transport anisotropy in de-twinned crystals[4] and the suppression of ortho-rhombicity at the superconducting transition.[5] In the over-doped region, the superconductivity is seen in the tetragonal structure. [1,3]

Ever since their discovery it has been realized that the FeAs based superconductors have a large upper critical field.[6] Experiments using high magnetic fields facilities have suggested that the upper critical field can show paramagnetic limited behaviour.[7] Evaluation of $H_{C2}$ anisotropy ($\gamma$) of a superconductor is very important from both fundamental physics and application point of view. In $Ba_{1-x}K_xFe_2As_2$ [6] it has been suggested that the large upper critical field is associated with a low anisotropy, indicating the suitability of these compounds for applications. With the availability of Co substituted single crystals of $BaFe_2As_2$, the doping dependence of $H_{C2}$ anisotropy has been addressed.[3,8,9] The reports are in general agreement with one another and suggests that $\gamma$ is an increasing function of temperature. A peaking of $\gamma$ value close to $T_C$ has been emphasized in some studies.[9] This unusual temperature dependence of $\gamma$ requires verification by more experiments. In particular it would be interesting to evaluate the role of Co stoichiometry on the $\gamma$ variation.

In this paper, we report on careful upper critical field anisotropy measurements on three Co substituted single crystals belonging to the under-doped and over-doped regimes of the phase diagram. We observe an increase in $\gamma$ with increase in Co content. $\gamma$ also increases with temperature for all Co compositions investigated and shows a clear evidence for a peak close to $T_C$. We also make an attempt to understand the $H_{C2}$ anisotropy based on recent models. [10-13] In addition this paper addresses the magneto-resistance anisotropy at the SDW anomaly in the under-doped compound. All the single crystals studied in the present work were synthesized without any flux.

The $BaFe_{2-x}Co_xAs_2$ ($x_{nominal}$ = 0.10, 0.15 and 0.20) single crystals were prepared using stoichiometric mixtures of Ba chunks and FeAs and CoAs powders. The samples were heated, at 50 °C/hour, held at 1190 °C for 24 hours, and then slow cooled to 800 °C at a rate of 2 °C/hour, which was followed by a fast rate of cooling of 50 °C/hour upto room temperature. Large number of small shiny crystals, with flat plate like morphology were found in the crucibles. Some of the crystals have sizes of upto 7 mm × 3 mm × 0.4 mm. These plate-like single crystals were characterized by X-ray Diffraction (XRD), energy dispersive X-ray (EDX) spectroscopy, DC resistivity and magneto-resistance (MR) measurements. Phase purity of the crystals were determined by powder XRD on ground single crystals. The XRD patterns showed peaks corresponding to the tetragonal space group symmetry I4/mmm (Space group No: 139) of $BaFe_2As_2$, without any impurity phases. XRD pattern taken on cleaved pieces of the crystals showed only (*00l*) diffraction peak, suggestive of the crystallographic *c*-axis orientation being perpendicular to the flat surface of the crystal. The lattice parameters were calculated from powder XRD data, for the tetragonal crystal structure of space group I4/mmm and their variation as a function of Co concentrations are given in Table I. The values for lattice constants are consistent with the reported data for the $BaFe_{2-x}Co_xAs_2$ systems.[3] Based on the reported unit cell volume and Co composition correlation,[3] the amount of Co present in the crystals were estimated. The average value Co fraction $x_{av}$ shown in Table I, is estimated from the cell volume and EDX based results, which will henceforth be used to identify the crystals in the manuscript.

TABLE I: Properties of the $BaFe_{2-x}Co_xAs_2$ single crystals studied in the present work. $x_V$ and $x_{EDX}$ are the estimated Co content based on unit cell volume data reported in Ni *et. al.* [3] and EDX results. $x_{av}$ is the average value of $x_V$ and $x_{EDX}$.

| $x_{nominal}$ | *a* axis Å | *c* axis Å | V Å$^3$ | $x_V$ | $x_{EDX}$ | $x_{av}$ |
|---|---|---|---|---|---|---|
| 0.10 | 3.9603 | 12.999 | 203.89 | 0.079 | 0.084 | 0.082 |
| 0.15 | 3.9600 | 12.993 | 203.74 | 0.109 | 0.125 | 0.117 |
| 0.20 | 3.9596 | 12.984 | 203.56 | 0.145 | 0.141 | 0.143 |

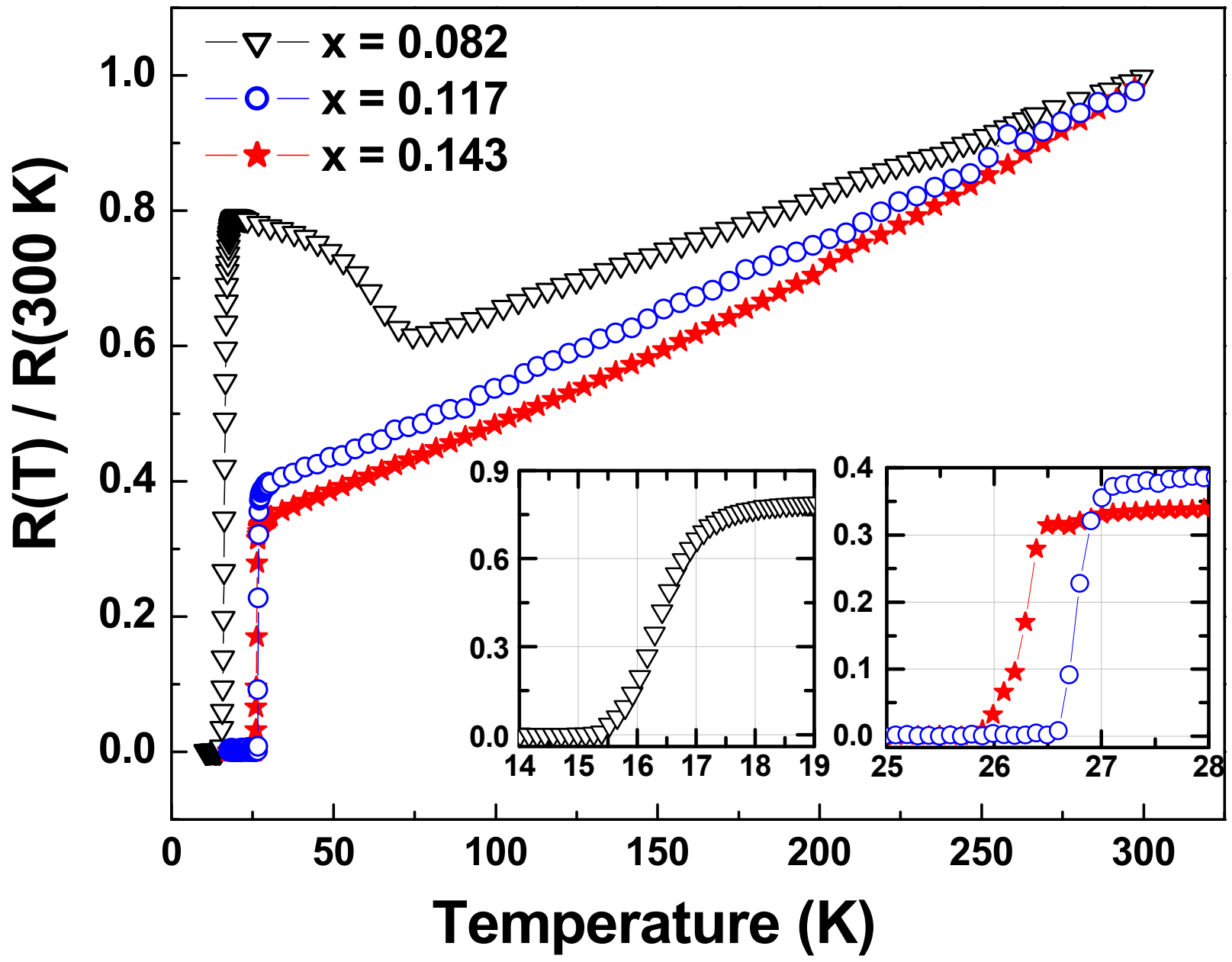

FIG. 1: (Color online) Temperature dependence of normalized electrical resistance of the $BaFe_{2-x}Co_xAs_2$ single crystals. Insets show enlarged view near the superconducting transition.

Figure 1 shows the variation of normalized electrical resistance with temperature for all the single crystals, measured on the *ab* planes. For the *x* = 0.117 and *x* = 0.143 crystals, normal state resistance decreased monotonically before dropping to zero at $T_C$. On the other hand, the normal state resistance for the *x* = 0.082 sample shows decrease upto ~ 75 K, below which it increases prior to the precipitous fall at the superconducting $T_C$. This behavior of normal state resistivity is thought to be associated with the structural and spin density wave transitions for this Co fraction.[1,3] The inset of Fig. 1 shows a blow-up around the superconducting transition. The values of $T_C$ and $\Delta T_C$ obtained from Fig. 1 are given in Table II. The $T_C$ is determined by the 90% criterion, viz., the temperature corresponding to the point where the resistivity/resistance falls to 90% of its normal state value and $\Delta T_C$ is determined using the difference of $T_{90\%}$ and $T_{10\%}$. It is to be remarked that $T_C$'s in the present single crystals are higher by ~ 3 - 4 K as compared to most of the reported $T_C$ values for $BaFe_{2-x}Co_xAs_2$ system.[1,3] Consistent results were obtained for measurements on 5 different pieces from each composition with scatter on $T_C$ limited to 0.5 K. Such increased $T_C$ has been reported for crystals grown by flux method also. [14]

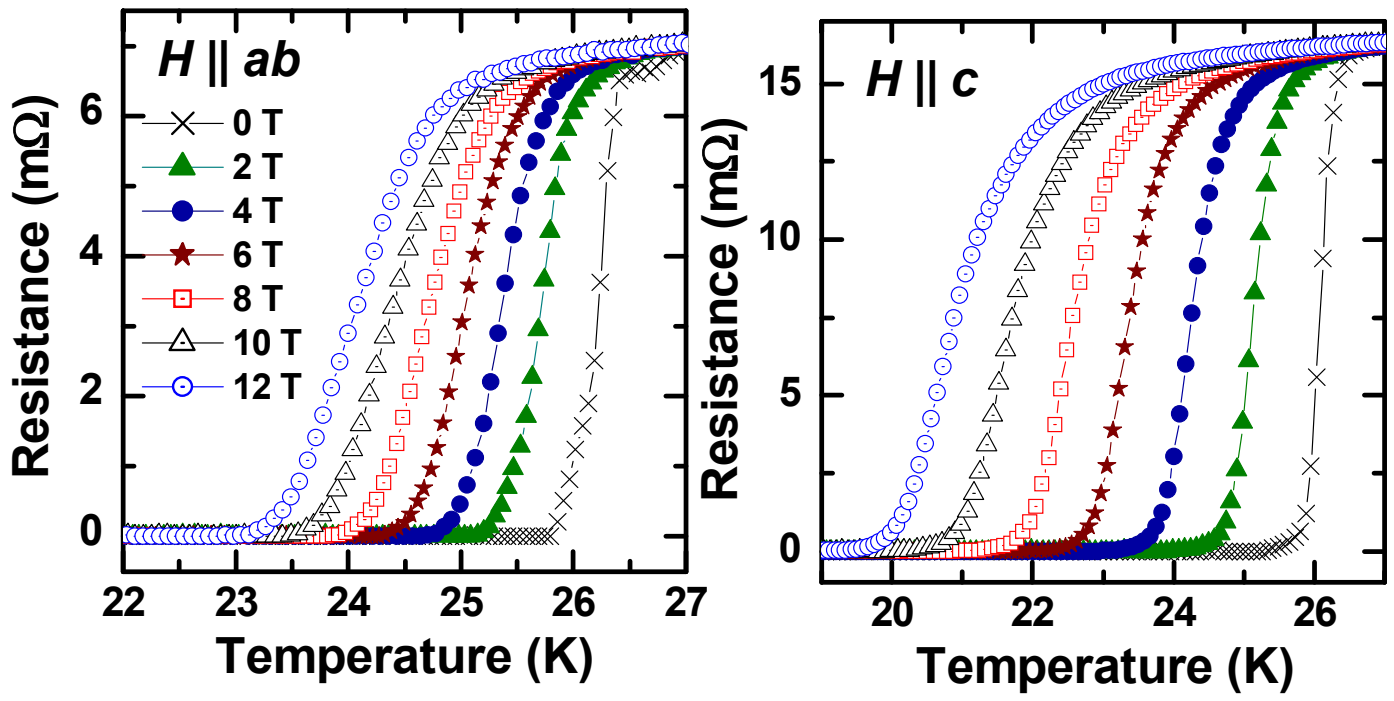

FIG. 2: (Color online) Temperature dependence of resistance for H || ab (left) and H || c (right) configurations for $x = 0.143$ sample, for external magnetic fields indicated.

Upper critical field behavior as a function temperature was investigated by magneto transport measurements, upto 12 T with field applied parallel and perpendicular to the *ab* plane. Figure 2 shows the resistance versus temperature for the sample, $x = 0.143$, for representative fields, in the two directions, in separate panels. Decrease in $T_C$ with increase in field is evident from Fig. 2, and the decrease is larger for H || c. Figure 3 shows the variation of $H_{C2}$ as a function of the reduced temperature, $t = T/T_C$ for the three Co concentrations, for H || ab and H || c. For all the Co concentrations and for both H || ab and H || c, $H_{C2}$ exhibits almost linear temperature dependence near $T_C$. $[dH_{C2}/dT]_{TC}$ obtained from the data are listed in Table II and values are in agreement with earlier reports. [3,8,9]

Recently Gurevich [10] showed that, in the arsenide superconductors, where inter-band scattering is dominant, the high temperature $H_{C2}$ has two limiting forms viz. $H_{C2}(t) \propto (1-t)$, in the weak paramagnetic regime and $H_{C2}(t) \propto \sqrt{(1-t)}$, for the paramagnetic limited region. [10] The present data fits well with $H_{C2}(t) \propto (1-t)$ for both H || ab and H || c configurations and for the three Co concentrations investigated. Figure 3 also shows expected behaviour for $H_{C2}(t) \propto \sqrt{(1-t)}$. According to Gurevich,[10] $H_{C2}$ in arsenides is sensitive to band parameters and in particular $[1/(T_C)]([(dH_{C2})/dT])_{TC} \propto [1/(c_1v^2{}_1 + c_2v^2{}_2)]$ where $c_1$ and $c_2$ are functions of the inter and intra-band coupling constants and $v_1$ and $v_2$ depends on the Fermi velocity in the two bands. In the case of iron arsenides since inter-band coupling is stronger compared with the intra-band coupling, $c_1 \approx c_2$, and $[1/(T_C)]([(dH_{C2})/dT])_{TC} \propto (v^2{}_1+v^2{}_2)^{-1}$.

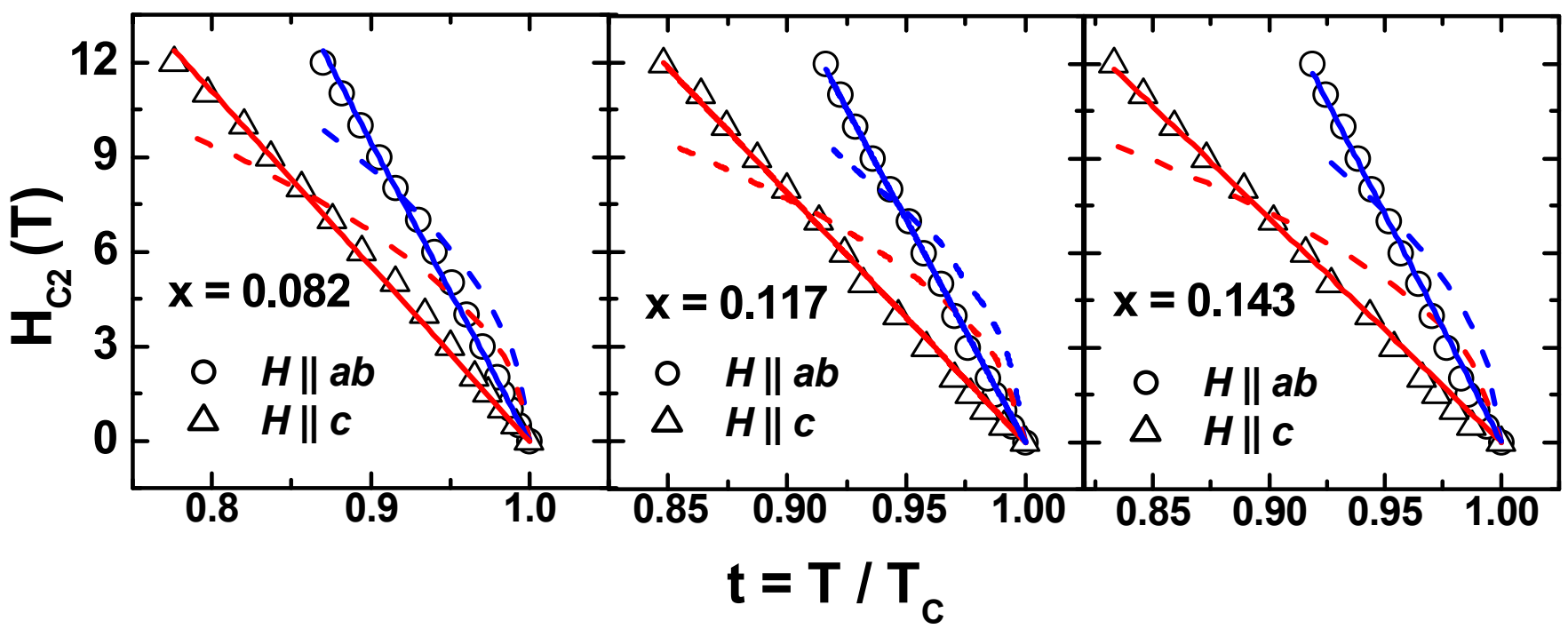

FIG. 3: (Color online) Upper critical fields of the $BaFe_{2-x}Co_xAs_2$ single crystals, as a function of reduced temperature. Circles and triangles shows experimental values for H || ab and H || c configurations. Solid and dashed lines (blue color: H || ab and red color: H || c) are fit to $H_{C2} \propto (1-t)$ and $H_{C2} \propto \sqrt{(1-t)}$ respectively.

The calculated values of $[1/(T_C)]([(dH_{C2})/dT])_{TC}$ for the three Co concentrations are presented in Fig. 4. The values obtained from the data of Ni *et. al.* [3] and Kano *et. al.* [9] are also shown in the figure for comparison. An overall agreement between the $[1/(T_C)]([(dH_{C2})/dT])_{TC}$ values measured from different groups is evident. In general, $[1/(T_C)]([(dH_{C2})/dT])_{TC}$ is lower for higher Co concentration. $[1/(T_C)]([(dH_{C2})/dT])_{TC}$ will be determined by the Fermi velocities of the electron-like and hole-like bands viz., $v^2{}_e+v^2{}_h$. It is known from LDA calculations [15,16] and ARPES [17,18] experiments that the hole Fermi surface shrinks and electron Fermi surface enlarges with increasing Co content and effective mass of electrons is lower than that of the holes, resulting in $v^2_e+v^2_h$ being dominated by $v^2_e$. The increase in $v^2_e$ with Co substitution can therefore give rise to the observed decrease of $[1/(T_C)]([(dH_{C2})/dT])_{TC}$ seen in Fig. 4.

TABLE II: Superconducting parameters of the $BaFe_{2-x}Co_xAs_2$ single crystals studied. Anisotropy of upper critical field determined by $\gamma = H_{C2} \parallel ab / H_{C2} \parallel c$.

| $x_{av}$ | $T_C$ (K) | $\Delta T_C$ (K) | $[(dH_{C2})/dT]_{TC}$ (T/K) H ‖ ab | H ‖ c | Anisotropy ($\gamma$) at $T/T_C = 0.95$ |
|---|---|---|---|---|---|
| 0.082 | 17.1 | 1.1 | 5.3 | 3.3 | 1.74 |
| 0.117 | 26.9 | 0.4 | 5.4 | 3.1 | 1.92 |
| 0.143 | 26.4 | 0.4 | 5.7 | 2.8 | 2.18 |

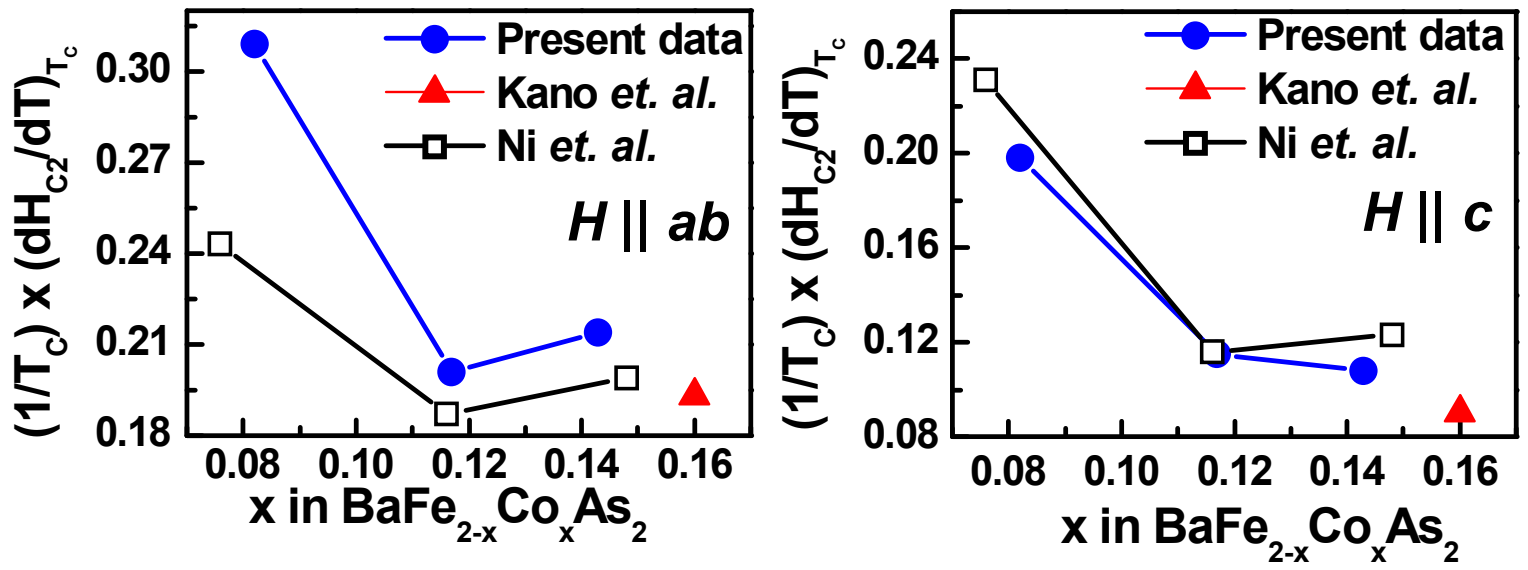

FIG. 4: (Color online) $[1/(T_C)]([(dH_{C2})/dT])_{TC}$ as a function of Co concentration, for H ‖ ab and H ‖ c configurations. Circles shows data from present work and squares and triangles are based on the data of Ni *et. al.* [3] and Kano *et. al.*. [9]

Figure 5 shows the anisotropy of the upper critical field, defined as $\gamma = H_{C2} \parallel ab / H_{C2} \parallel c$ as a function of temperature for the different Co concentrations. It is clear from the Fig. 5 that $\gamma$ increases with Co concentration. The values of $\gamma$ observed in the Co substituted samples are similar to that observed for the $Ba_{1-x}K_xFe_2As_2$ and $BaFe_{2-x}Co_xAs_2$ systems, [3,19,20] and the values of $\gamma$ measured at the reduced temperature of 0.95, are given in Table II. It has been shown by band structure calculations that in Co doped $BaFe_2As_2$ superconductor the hole Fermi surface strongly disperses along the crystallographic *c* direction.[18] Accordingly it is expected that the $H_{c2}$ anisotropy should only decrease with Co doping, which is contrary to the observation. The increase in γ with increase Co content might arise due to an increase in the ratio of inter/intra-band coupling. The importance of scattering in destabilizing the SDW state leading to the superconducting state has been suggested in recent calculations.[21]

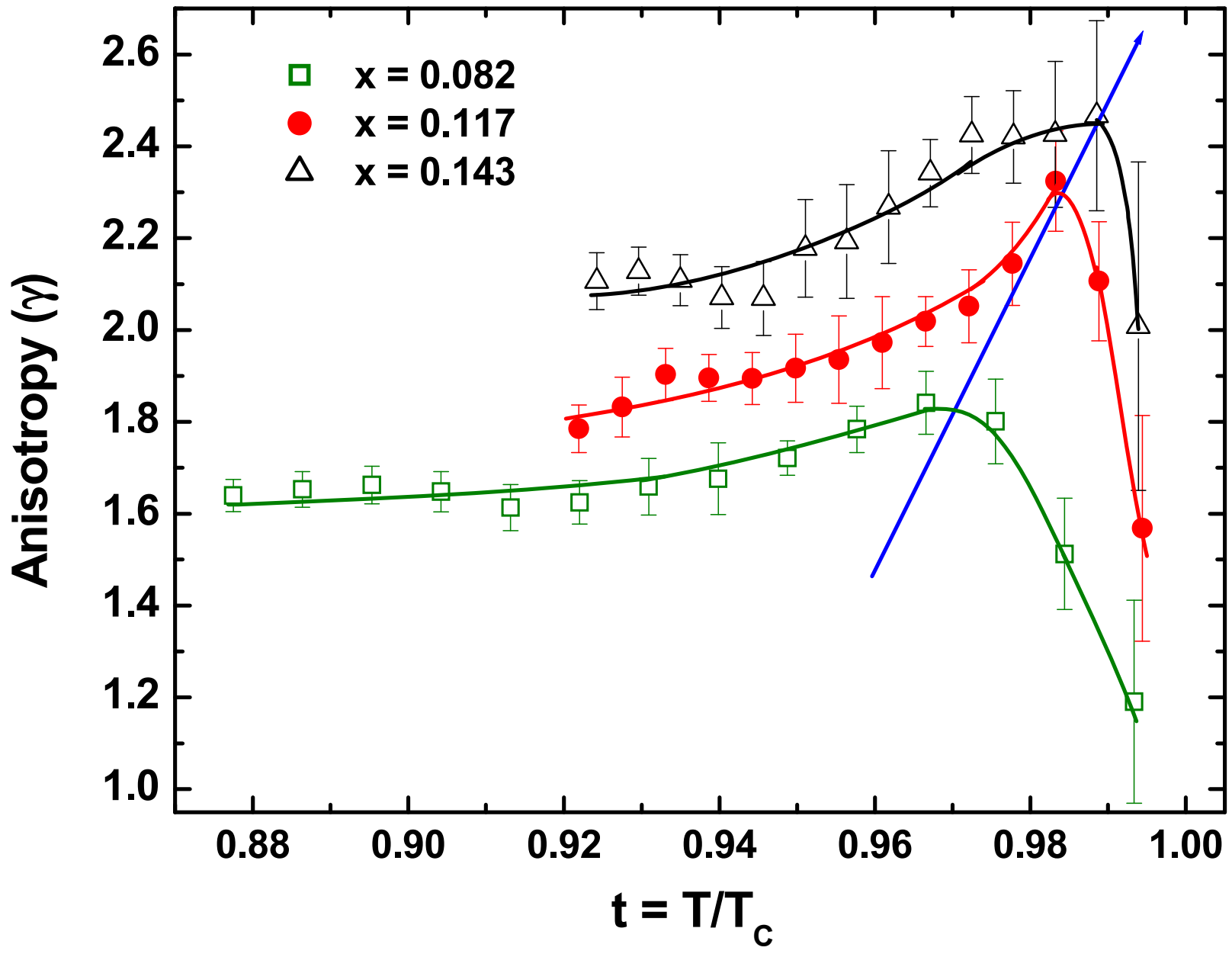

FIG. 5: (Color online) Anisotropy parameter (defined as, $\gamma = H_{C2} \parallel ab / H_{C2} \parallel c$) for the $BaFe_{2-x}Co_xAs_2$ single crystals. Data points are shown in symbols and the lines are guidelines to the eye. Arrow indicates the shift in peak position with Co concentration.

Figure 5 also indicates that the value of γ is temperature depended for the three Co concentrations. As the temperature is lowered from $T_C$, γ increases, passes through a peak and then decreases with further decrease in temperature, indicating more isotropic behavior at lower temperatures. This trend is same for the three Co concentrations studied. Also, the peak in γ is seen to shift to higher temperature with increase in Co concentration. The peak like anomaly observed in Fig. 5, close $T_C$ has been seen in only two of the earlier reports. [9,19]

In an isotropic superconductor, $\gamma$ is independent of temperature [11], whereas $\gamma$ in a two band system, it is expected to be temperature dependent. In the extensively studied two gap superconductor, $MgB_2$, $\gamma$ versus temperature (T) is a decreasing function of T, in clean single crystals [12,13] whereas in irradiated samples $\gamma$ is an increasing function T. [12,13] Calculations done for $MgB_2$ indicates that the temperature dependence of $\gamma$ depends on the relative diffusivities in the two bands [12,13], which can be altered by substitution and introduction of disorder. An increase in $\gamma$ with T, the observation of peak in $\gamma$ close to $T_C$ and it's shift to higher $T/T_C$ with increase in Co concentration are unmistakable in Fig. 5. These results contrast with that observed in $MgB_2$ and suggest the need for theoretical calculations of magnetic anisotropy versus temperature, in $BaFe_{2-x}Co_xAs_2$ system, taking into account inter-band scattering effects.

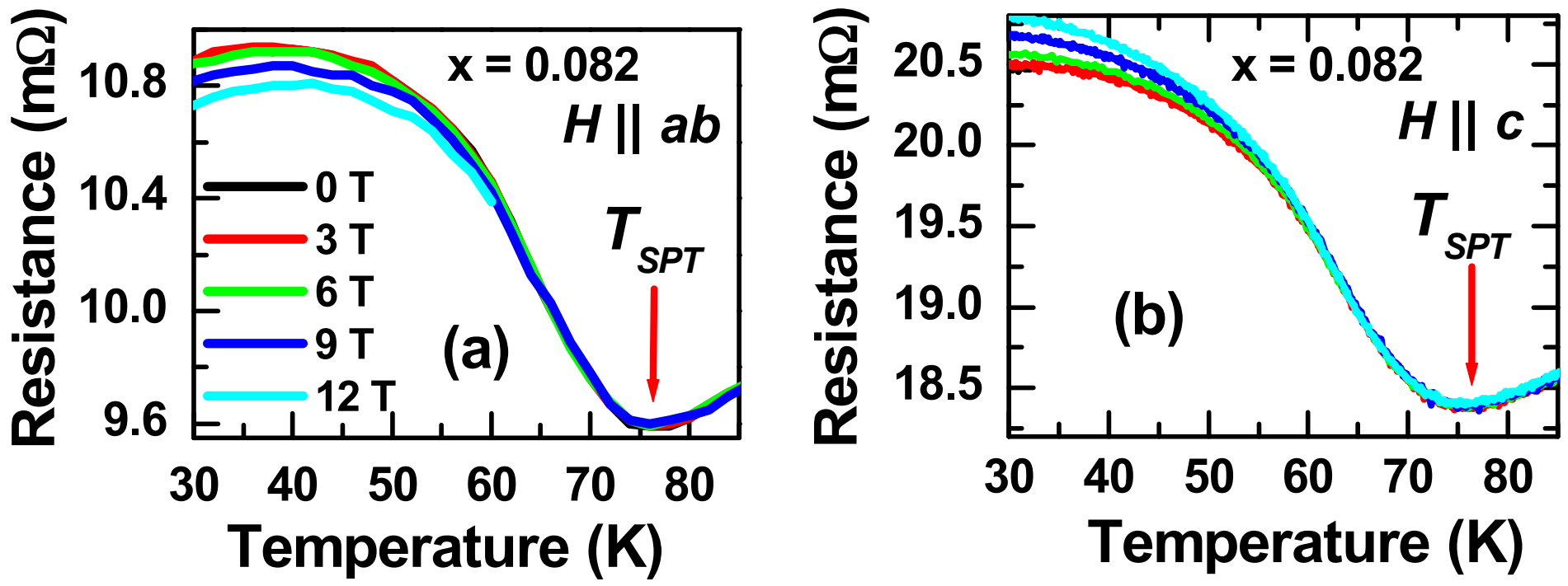

FIG. 6: (Color online) Temperature dependence of resistance in the SDW state for $x = 0.082$, for fields applied (a) H || ab and (b) H || c configurations. Arrow shows the structural transition temperature, $T_{SPT}$.

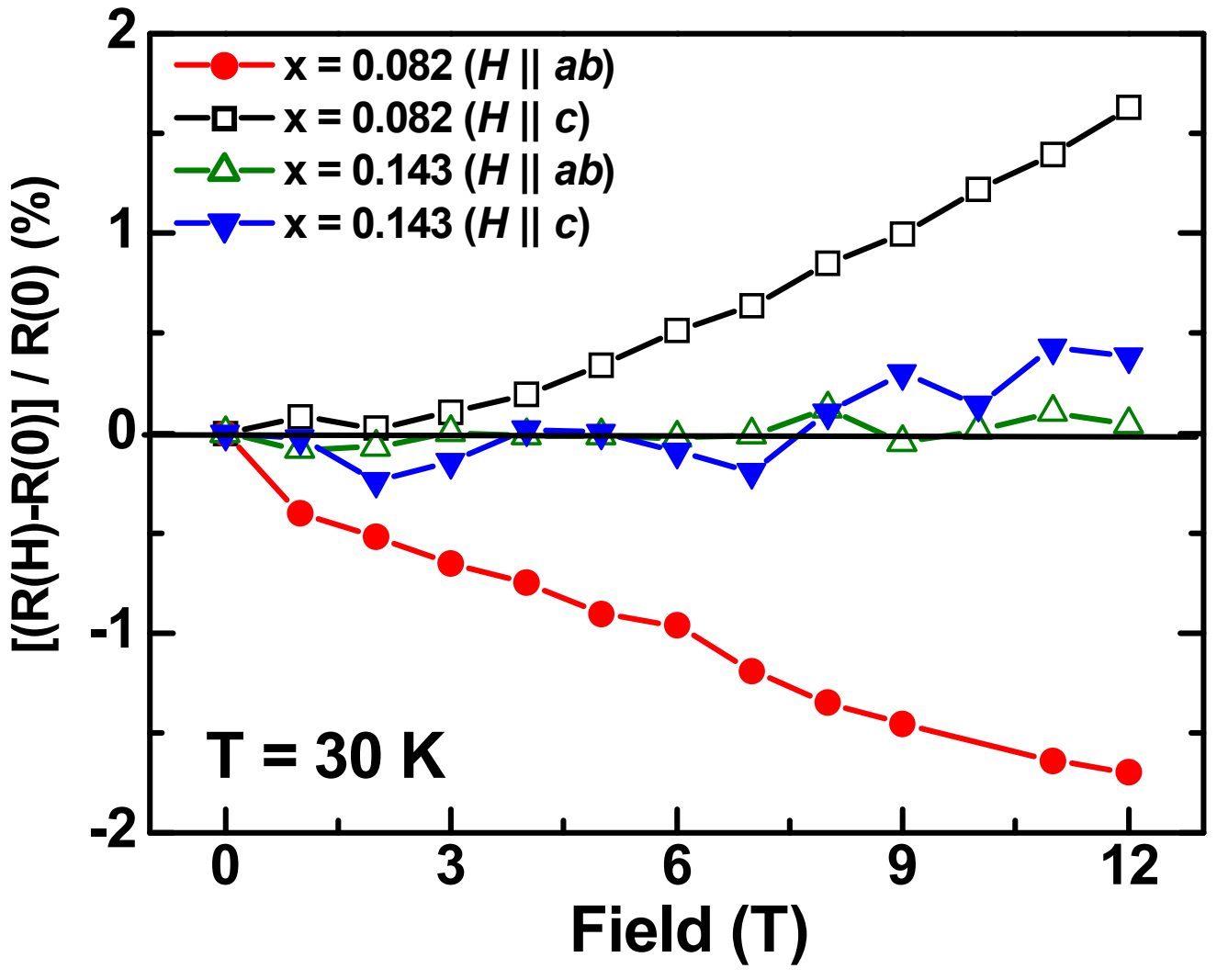

FIG. 7: (Color online) Magneto resistance in percentage for $x$ = 0.082 and $x$ = 0.143 samples, for 30 K.

We now turn our attention to the magneto-resistance in the SDW state for the $x$ = 0.082 crystal. Figure 6(a) and 6(b) shows the MR in the SDW state for the $x$ = 0.082, for fields applied H || ab and H || c. It is evident that only in the SDW state the crystal shows a noticeable negative MR for H || ab and positive MR for H || c. As the temperature increases closer to the structural transition temperature (~ 75 K) the MR is suppressed for both H || ab and H || c directions. Figure 7 compares the isothermal MR at 30 K for the x=0.082 for both H || ab and H || c directions. The figure clearly shows anisotropic MR for the sample. In order to compare this with the behaviour seen in overdoped sample, in Figure 7 is shown the MR data for the $x$ = 0.143 sample. It is evident that there is no noticeable MR in the normal state for the x=0.143 crystals as was also the case in the x=0.117 sample (not shown). The observed anisotropic MR shown in Fig. 6 and Fig. 7 is unusual when compared with the positive MR seen in undoped and K doped $BaFe_2As_2$ and in $SrFe_2As_2$.[22-24]

Summarizing, the behaviour of $H_{C2}$ and its anisotropy of $BaFe_{2-x}Co_xAs_2$ superconductors is studied using single crystals synthesized without flux. The $H_{C2}(t)$ is found to proportional to $(1-t)$ for both H || ab and H || c configurations. Anisotropy ($\gamma$) is found to increase with increase in Co content, and the anisotropy parameter shows a peak close to the $T_C$, which shifts to higher

T/Tc with increase in Co content. MR measurements in the $x$ = 0.082 crystal also show a significant anisotropy in the SDW region.

Vinod K acknowledges Department of Atomic Energy for the K.S. Krishnan Research Associateship. Authors acknowledge Ms S. Kalavathi for the XRD characterization and Sophisticated Analytical Instrumentation Facility, Indian Institute of Technology, Madras, for the EDX measurements.

**References**

[1] J. H. Chu, J. G. Analytis, C. Kucharczyk, and I. R. Fisher, Phys. Rev. B **79**, 014506 (2009).

[2] P. C. Canfield and S. L. Bud'ko, Annu. Rev. Condens. Matter Phys. **1**, 27 (2010).

[3] N. Ni, M. E. Tillman, J. Q. Yan, A. Kracher, S. T. Hannahs, S. L. Bud'ko, and P. C. Canfield, Phys. Rev. B **78**, 214515 (2008).

[4] J. H. Chu, J. G. Analytis, K. De Greve, P. L. McMahon, Z. Islam, Y. Yamamoto, and I. R. Fisher, Science **329**, 824 (2010).

[5] S. Nandi, M. G. Kim, A. Kreyssig, R. M. Fernandes, D. K. Pratt, A. Thaler, N. Ni, S. L. Bud'ko, P. C. Canfield, J. Schmalian, R. J. McQueeney, and A. I. Goldman, Phys. Rev. Lett. **104**, 057006 (2010).

[6] H. Q. Yuan, J. Singleton, F. F. Balakirev, S. A. Baily, G. F. Chen, J. L. Luo, and N. L. Wang, Nature **457,** 565 (2009).

[7] G. Fuchs, S. L. Drechler, N. Kozlova, M. Bartkowiak, J. E. Hamann-Borrero, G. Behr, K. Nenkov, H. H. Klauss, H. Maeter, A. Amato, H. Luetkens, A. Kwadrin, R. Khasanov, J. Freudenberger, A. Köhler, M. Knupfer, E. Arushanov, H. Rosner, B. Büchner, and L. Schultz, New J. Phys. **11,** 075007 (2009).

[8] A. Yamamoto, J. JaroszynskiJ, C. Tarantini, L. Balicas, J. Jiang, A. Gurevich, D. C. Larbalestier, R. Jin, A. S. Sefat, M. A. McGuire, B. C. Sales, D. K. Christen, and D. Mandrus, Appl. Phys. Lett. **94**, 062511 (2009).

[9] M. Kano, Y. Kohama, D. Graf, F. Balakirev, A. S. Sefat, M. A. Mcguire, B. C. Sales, D. Mandrus, and S. W. Tozer, J. Phys. Soc. Jpn. **78**, 084719 (2009).

[10] Gurevich, Phys. Rev. B **82**, 184504 (2010).

[11] V. G. Kogan, S. L. Bud'ko, Physica C **385**, 131 (2003).

[12] A. Golubov and A. E. Koshelev, Phys. Rev. B **68**, 104503 (2003).

[13] A. Gurevich, Phys. Rev. B **67**, 184515 (2003).

[14] Y. Nakajima, T. Taen, and T. Tamegai, J. Phys. Soc. Jpn. **78**, 023702 (2009).

[15] A. S. Sefat, R. Jin, M. A. McGuire, B. C. Sales, D. J. Singh, and D. Mandrus, Phys. Rev. Lett. **101**, 117004 (2008).

[16] D. C. Johnston, Advances in Physics **59**, 803 (2010).

[17] P. Vilmercati, A. Fedorov, I. Vobornik, U. Manju, G. Panaccione, A. Goldoni, A. S. Sefat, M. A. McGuire, B. C. Sales, R. Jin, D. Mandrus, D. J. Singh, and N. Mannella, Phys. Rev. B **79**, 220503(R) (2009).

[18] S. Thirupathaiah, S. de Jong, R. Ovsyannikov, H. A. Dürr, A. Varykhalov, R. Follath, Y. Huang, R. Huisman, M. S. Golden, Y. Z. Zhang, H. O. Jeschke, R. Valentí, A. Erb, A. Gloskovskii, and J. Fink, Phys. Rev. B **81**, 104512 (2010).

[19] Z. S. Wang, H. Q. Luo, C. Ren, and H. H. Wen, Phys. Rev. B **78**, 140501(R) (2008).

[20] M. A. Tanatar, N. Ni, C. Martin, R. T. Gordon, H. Kim, V. G. Kogan, G. D. Samolyuk, S. L. Bud'ko, P. C. Canfield, and R. Prozorov, Phys. Rev. B **79**, 094507 (2009).

[21] H. Wadati, I. Elfimov, and G. A. Sawatzky, Phys. Rev. Lett. **105**, 157004 (2010).

[22] G. F. Chen, Z. Li, J. Dong, G. Li, W. Z. Hu, X. D. Zhang, X. H. Song, P. Zheng, N. L. Wang, and J. L. Luo, Phys. Rev. B **78**, 224512 (2008).

[23] X. F. Wang, T. Wu, G. Wu, H. Chen, Y. L. Xie, J. J. Ying, Y. J. Yan, R. H. Liu, and X. H. Chen, Phys. Rev. Lett. **102**, 117005 (2009).

[24] H. Q. Luo, P. Cheng, Z. S. Wang, H. Yang, Y. Jia, L. Fang, C. Ren, L. Shan, and H. H. Wen, Physica C **469**, 477 (2009).